\documentclass{aa}
\usepackage{psfig}

\def \rsun {\ifmmode$R$_{\odot}\else R$_{\odot}$\fi}

\def \hcm {\hbox {\ifmmode $ H atoms cm$^{-2}\else H atoms cm$^{-2}$\fi}}
\def\approxgt{\mathrel{\hbox{\rlap{\lower.55ex \hbox {$\sim$}}
        \kern-.3em \raise.4ex \hbox{$>$}}}}
\def\approxlt{\mathrel{\hbox{\rlap{\lower.55ex \hbox {$\sim$}}
        \kern-.3em \raise.4ex \hbox{$<$}}}}
\newcommand {\sax} {{BeppoSAX}}

\begin{document}

\thesaurus{06 (08.09.2: $\gamma$ Cas; 08.05.2; 08.23.1; 13.25.5)}


\title{BeppoSAX broad--band observations of Gamma Cassiopeiae}

\author{Alan Owens\inst{1}, 
T. Oosterbroek\inst{1},
A.N. Parmar\inst{1},
R. Schulz\inst{2},
J.A. St\"uwe \inst{3},
F. Haberl \inst{4}
}

\institute{
Astrophysics Division, Space Science Department of ESA,
ESTEC, 2200 AG Noordwijk, The Netherlands
\and
Solar System Division,  Space Science Department of ESA,
ESTEC, 2200 AG Noordwijk, The Netherlands
\and
Astronomisches Institut der Ruhr-Universit\"at Bochum, Germany
\and
Max-Planck-Institut fur Extraterrestrische Physik,
Giessenbachstra{\ss}e, D-85748, Garching, Germany
}

\offprints{Alan Owens (aowens@astro.estec. esa.nl)}

\date{Received 11 February 1999; accepted 19 April 1999 }

\authorrunning{A. Owens et al.}

\maketitle

\begin{abstract}
We report broad-band X-ray measurements of the Be star $\gamma$ Cassiopeiae 
by the BeppoSAX X-ray astronomy satellite. The observations took place on 
1998 July, 18--23. The 0.1--200~keV X-ray spectrum is reasonably well fit 
by an optically thin thermal plasma model of temperature 12.5$\pm$0.6~keV 
with significant residuals around 0.3~keV and 1~keV. The former is 
interpreted as the variable soft component reported by ROSAT, although 
there is no evidence for variability at the 5\% level. For a blackbody 
interpretation, the fitted temperature is 100$\pm ^{320} _{13}$~eV, in 
agreement with the ROSAT value of 200$\pm$ 10~eV. However, a MEKAL 
interpretation gives a significantly lower temperature of (48$\pm$11~eV). 
The fitted abundances are about half solar values, in agreement with 
previous measurements. At higher energies, the spectrum does not require 
non-thermal components and the observation of a line at 6.8 keV supports 
the ASCA interpretation of the source as an accreting white dwarf.
Assuming a source distance of 188~pc, the bolometric luminosity in the 
2--10~keV band is 6$\times$10$^{32}$~ergs~s$^{-1}$. Simultaneous optical 
measurements by the Wendelstein Observatory near Munich, indicate 
that the source continues to be in a late but rather normal Be phase, 
with no obvious signs of a transition to the Be-shell phase. The measured 
magnitudes at B, V and R wavelengths of 2.18$\pm$0.06, 2.23$\pm$0.02 and 
2.36$\pm$0.03, respectively, confirm this.

\end{abstract}


\section{Introduction}
Gamma Cassiopeiae is a B0.5 IVe star which has been extensively
studied at optical wavelengths. 
It is also a moderately strong X-ray source ($\sim$10$^{33}$erg~s$^{-1}$) 
and is believed to form part of a binary system, based on its 
similarities to the Be X-ray binary, X Persei (Mason et al. 1976). 
However, despite two decades of intense 
study, its status as an X-ray binary remains unclear. 

 
The phenomenological model of $\gamma$--Cas has been derived largely from
optical interferometric and spectropolarmetric measurements (e.g., 
Horaguchi et al. 1994; Stee et al. 1995; Quirrenbach et al. 1997). They 
reveal a pronounced mass loss from the system (which implies that material 
is available for accretion onto a companion) and the presence of a
substantial circumstellar envelope, which in turn, may provide the source of
material for a companion to accrete from (Vakili et al. 1994). The
presence of a companion is inferred from the observed 
oscillations in the ratio of V/R peak intensities in double peaked
H$_\alpha$ line profiles. Additional evidence is provided by an 
analysis of a 20~ks EXOSAT observation by Frontera et al. (1987). 
In the 2--8~keV energy range, the observed X-ray flux is 
$\sim$2 orders of magnitude greater than that expected for 
a star with spectral class earlier than A (Pallavicini et al. 1981) 
and the 2--20~keV spectrum 
is consistent with that commonly observed in low luminosity Be/neutron binary star 
systems. The derived spectral index is $\sim$1.5 for a power-law 
interpretation, or alternately, a temperature of $\sim$10--15~keV for 
optically thin bremsstrahlung emission. A timing analysis 
of the data failed to shed any insight into the orbital dynamics 
of the system, but did suggest a possible period of $\sim$100 
mins. However, Parmar et al. (1997) found no evidence for this 
periodicity in a subsequent 30~hr continuous EXOSAT observation, suggesting instead 
that the apparent periodicity is produced by statistical fluctuations. 
In common with many other X-ray binaries, chaotic 
X-ray variability with no preferred timescale down to $\sim$10~s has 
also been observed (Kubo et al. 1988) and is thought to arise from 
fluctuations in the rate of accretion onto the compact object.
 
\subsection{Neutron star, white dwarf or coronal emission?}
Its X-ray luminosity of 10$^{33}$ ergs s$^{-1}$ is consistent
with both coronal emission (as observed from OB stars) and wind powered 
emission from either a neutron star or white dwarf. 
However, the measured temperature of $\sim$12~keV 
is much hotter than found in OB stars, which usually have 
temperatures in the range
$\sim$0.5--2.5 keV. 
Evidence that the compact companion in the $\gamma$--Cas 
system is actually a white dwarf, rather than a neutron star, is 
strongly suggested by the {\it Tenma} observation of strong Fe line emission
at 6.8~keV of equivalent width, EW, 280~eV (Murakami et al. 1986).
This implies a highly ionized plasma of temperature, $kT$ $\sim$10~keV which 
is commonly seen in white dwarf systems (Mukai \& Shiokawa 1993).  
High mass neutron star accreting systems, on the other hand, 
usually show Fe emission at 6.4~keV, with occasionally weak
(EW$<100$~eV) emission at 6.7~keV.  An 8.4~ks ROSAT PSPC observation of
$\gamma$--Cas (Haberl 1995) revealed evidence for a soft component which
can be modeled as black body radiation of kT$\sim$200~eV. It is believed
to emanate from the heated surface of a white dwarf near the magnetic pole.  
This component appears to be modulated with a period of 135.3 mins
which may be the spin period of the white dwarf. The modulation may 
then arise from geometrical self 
occultation by the white dwarf (King \& Shaviv 1984), or by photoelectric 
absorption, as in the accretion curtain model of Rosen et al. (1988). 
Based on a possible stellar modulation in the chaotic X-ray emission
and a similarity (spectrally and temporally) to late-type flaring stars 
(such as RS CVns), Smith et al. (1998) proposed a mechanism 
in which X-rays are produced in magnetically generated hot spots 
on the surface of $\gamma$--Cas itself. Such an origin would be consistent 
with the apparent lack of orbital motion and the active 
Be nature of the star.

In this paper we report on simultaneous optical and X-ray measurements 
with the BeppoSAX satellite and present the first 0.1--50~keV broad-band 
energy spectrum of $\gamma$--Cas.

\section{Optical observations}
As $\gamma$--Cas was one of the first stars observed to display the
characteristics nowadays categorized as Be star phenomenon, 
the collection of optical observations covers a very long time 
baseline (e.g., AFOEV 1998). Based on photometric and spectroscopic 
peculiarities in common with other Be stars, Doazan et al. (1983) 
describe a three phased ``cycle'' in the long term behavior of 
$\gamma$--Cas, without implying any periodicity. Such a cycle starts 
with the building-up of Balmer emission. This phase is characterized 
by moderate irregular variability in the Balmer line intensities as 
well as in the visual magnitude. The line intensities and magnitudes 
show a slow increase over several decades (Be phase). This behavior 
culminates in a second phase of high variability in both
characteristics (Be-shell phase) lasting approximately 10 years,
which was last observed in $\gamma$--Cas from 1932 to 1942. The cycle 
is terminated by a third phase with no detectable Be characteristics
(B-normal phase) lasting about 5 years. The visual magnitude of 
$\gamma$--Cas has more or less continuously risen from its early Be 
phase value of about 2\fm 6 in 1950 to a present value of $\sim$2\fm 2. 

Contemporaneous to the BeppoSAX observations, optical monitoring of 
$\gamma$--Cas was carried out with Universit\"ats-Sternwarte M\"unchen's 
80$\,$cm Telescope located at the Wendelstein Observatory. Measurements 
were made using the {\sc Monica} instrument in Johnson B, V and R around 
01:50 UT on 1998, July 21. The measured magnitudes were: B=2.18$\pm$0.06, 
V=2.23$\pm$0.02 and R=2.36$\pm$0.03, which confirms the ``normal'' behavior 
of $\gamma$--Cas at this time. Moreover, the resulting B--V=$-$0\fm 05 
is in perfect agreement with a correlation between visual magnitude 
and B-V color index described by Horaguchi et al. (1994),
which shows that, the brighter the V magnitude of $\gamma$--Cas,
the redder it becomes in B--V. This trend is commonly interpreted as 
being due to the growth of a circumstellar envelope with time.
In summary, $\gamma$--Cas, as indicated by its visual observables, 
is currently in a late but rather normal Be phase, with no obvious 
signs of a transition to the spectacular Be-shell phase.

\section{X-ray observations}
The X-ray observation was carried out using the BeppoSAX 
astronomy satellite (Boella et al. 1997a). The platform contains four 
coaligned Narrow Field Instruments (NFI) providing broad-band 
coverage over the energy range 0.1 to 300 keV. The NFI are: the 
Low Energy and Medium Energy Concentrator Spectrometers (LECS 
and MECS), the High Pressure Gas Scintillation Proportional Counter 
(HPGSPC) and the Phoswich Detection System (PDS). The LECS covers the
energy range 0.1--10~keV with an energy resolution of 8\,\% at 6~keV 
(Parmar et al. 1997). The MECS consists of three detectors similar to 
the LECS, with identical energy resolution but with thicker entrance 
windows, spanning 1.3--10~keV (Boella et al. 1997b). The HPGSPC 
is sensitive over the energy range 4--120~keV with an energy resolution 
of 4\,\% at 60~keV  (Manzo et al. 1997). The PDS covers the high energy 
range 13--300~keV with an energy resolution of 15\,\% at 60~keV (Frontera 
et al. 1997).
 
$\gamma$--Cas was observed from 1998 July 20 11:21 UTC to 
July 21 22:15 UTC, yielding total on-source exposure times 
of 10.2, 37.4, 20.0 and 18.6 ks in the LECS, MECS, HPGSPC and 
PDS, respectively.

\begin{figure}
\centerline{\psfig{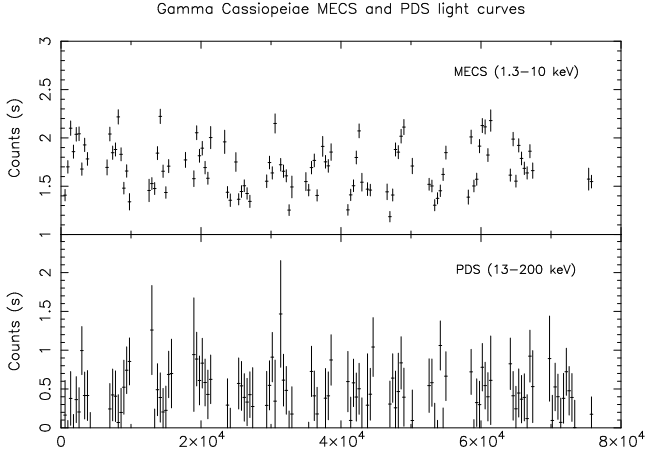}}
\caption[]{ The background subtracted 1.3-10 keV and 13-200 keV X-ray light curves of $\gamma$--Cas 
measured by the MECS and PDS. The data are binned in 400~s intervals. Considerable
variability is apparent in the MECS, even down to time scales of 10~s
when finer time resolution is used}
\end{figure}
 
\subsection{Data Analysis}
Data were processed with the SAXDAS data analysis system using
standard procedures. For the LECS and MECS, source events were 
extracted from regions of radii 8 and 4$'$, respectively, centered 
on the source position. Spectral fitting was carried out using 
{\sc XSPEC} (version 10.1). The LECS data were fit over the energy 
range 0.1 to 10~keV; the MECS 1.85 to 10~keV; the HPGSPC 7 to 65~keV 
and the PDS 13 to 200~keV. To ensure that the fitting statistic 
($\chi^2$) was unbiased across the entire energy range, data were 
rebinnned to a width of 33\% of the full width at half maximum 
energy resolution function of each instrument and also under the
condition that each bin contained a minimum of 20 counts. It is 
known from inter-instrument spectral calibrations that there can 
be small position dependent normalization differences between 
the instruments. Therefore, these factors were included as free 
multiplicative parameters during multiple NFI spectral fitting.

\begin{figure}
\centerline{\psfig{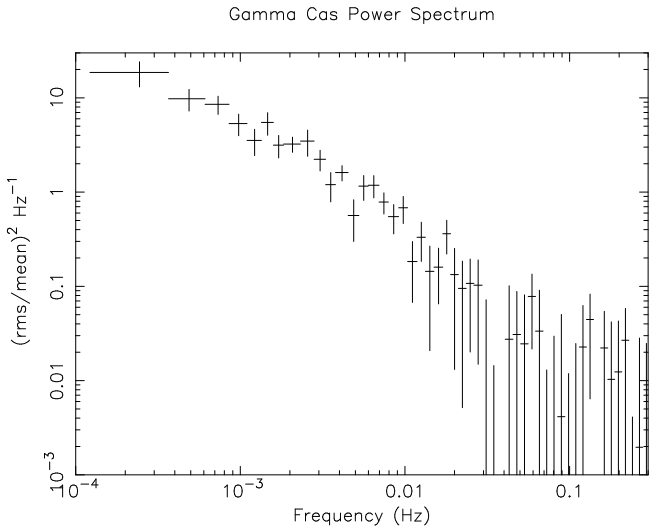}}
\caption[]{The 1.3--10~keV power density spectrum of $\gamma$-Cas. 
The Nyquist frequency corresponds to 0.5 Hz }
\end{figure}
 
\section{X-ray results}

\subsection{Intensity variations}

Fig. 1 shows low and high energy light curves measured by the MECS 
(1.3--10.0 keV) and the PDS (13--200~keV) after background subtraction. 
The time resolution is 400~s. The source intensity is clearly variable 
with a doubling of the flux occurring on the time scale of adjacent bins. The 
fastest observed fluctuation in the MECS is on a time scale of $\sim$10~s, 
consistent with the ASCA result of Kubo et al. (1998). In the PDS, the fastest 
observable variation is of the order of $\sim$40~s. This time scale 
implies an upper limit on the size of the emission region of a few 
$\times$ 10$^{11}$~cm. Next a search for periodic variations was carried out. 
On short time scales ($f \sim$10$^{-4}$--0.5~Hz), a power density curve 
reveals a ${1/f}$--type distribution, but no clear periodicities 
(see Fig. 2). At much longer periods, a period search reveals weak 
enhancements at 35 and 145 mins. We estimate 
that for narrow QPO/periodicities in the 0.01 Hz to 0.1 Hz range, we could detect 
a 10\% amplitude modulation at the 3$\sigma$ level.

In order to study the long-term X-ray variability of $\gamma$--Cas, 
we have also analyzed R-XTE All Sky Monitor (ASM) data which has 
continuously observed the source 
from 1996 February, 20 to 1998 December, 31. We have searched for periods in
the range 30--500 days using the Lomb-Scargle periodogram, using both
the individual dwell and 1-day averages data. No peaks with a high 
significance (i.e., $>$99\%) were found in the individual dwell data, 
although one peak at a period of $\sim$200 days was of marginal 
significance (at the $\sim$90\% level). As a check on its reality, we 
replaced the R-XTE ASM measurements with data drawn from a Gaussian 
distribution centered on zero with a $\sigma$ of 1 (i.e. no signal). 
We note that the peak still existed (albeit with a much lower significance), 
implying that it is probably caused by a windowing effect and therefore 
does not reflect a true period in the X-ray flux of $\gamma$--Cas. 
Analysis of the 1-day averaged data yielded similar results, i.e., no 
significant periodicities. We estimate that for periods around
100 days, we would have detected a 10\% periodic modulation at the 90\% confidence level.

 


\subsection{Spectral variations}
Simple models (i.e., power-laws, bremsstrahlung, etc.) gave poor fits
to the data. For example, an absorbed power-law gives a $\chi ^2$ of 
684 for 203 degrees of freedom (dof). The addition of an iron line 
at 6.77$\pm$0.03~keV improves the fit significantly ($\chi ^2$/dof=505/200),
but is still unacceptable at energies above 10~keV. Based on previous ROSAT 
and ASCA measurements (Haberl 1995; Kubo et al. 1998), we next 
investigated optically thin thermal plasma models with both 
the temperature and elemental abundance as free parameters. The results are listed in
Table 1. A best-fit MEKAL model, based in the calculations of Mewe and 
Kaastra (Mewe et al. 1986; Kaastra 1992) yields a $\chi ^2$/dof of 266/202 
for a fitted temperature of 12.5$\pm$0.7~keV and an abundance of 
0.43$\pm$0.05. These values are in agreement with the ASCA values of
10.7$\pm 0.6$~keV and 0.35$\pm$0.08 (Kubo et al. 1998). The fit
shows significant residuals around 300~eV and 1~keV. 
The addition of a carbon line at 277~eV results in a better fit 
($\chi ^2$/dof=244/199), which is significant under an $F$ test 
(P$>$99.9\%). If real, this could imply an overabundance of carbon. 
Following the ROSAT observation of a 
soft component, we next added a blackbody component. This resulted in a 
slightly worse fit ($\chi ^2$/dof=248/201) primarily due to residuals 
around 0.3~keV. The fitted temperature was 100$\pm ^{320} _{13}$~eV, which 
is consistent with the ROSAT value of 200$\pm$10~eV (Haberl 1995).
Assuming that the source is best described in terms of a multi--temperature
plasma, we replaced the blackbody 
component with a second, lower temperature, MEKAL component. The resulting 
$\chi ^2$/dof is 240/200 for temperatures 
of 12.3$\pm$0.6~keV and 0.05$\pm$0.01~keV. The fitted abundance is 
0.423$\pm $0.06, consistent with previous measurements. 
Whereas ROSAT reported marked variability in the soft component and suggested
a possible modulation period of 135 mins, our data are consistent with
a constant mean rate throughout the observation. For example, in the
0.1 to 0.5~keV band, a best--fit constant mean rate yields a $\chi ^2$/dof 
of 64/57, as compared to 423/64 for the 0.1--10~keV band. 
The combined NFI 
count spectrum and the best-fit two--temperature MEKAL model (folded 
through the 
appropriate instruments response functions) is shown in Fig. 4. In Fig. 5 
we show the corresponding best-fit incident photon spectrum. The observed flux 
in the 2--10~keV band is 1.33$\times$10$^{-10}$ erg~cm$^{-2}$ s$^{-1}$, which 
for a source distance of 188~pc (ESA 1997), corresponds to a 
bolometric luminosity of 5.7$\times$10$^{32}$~erg~s$^{-1}$.

\begin{figure}
\centerline{\psfig{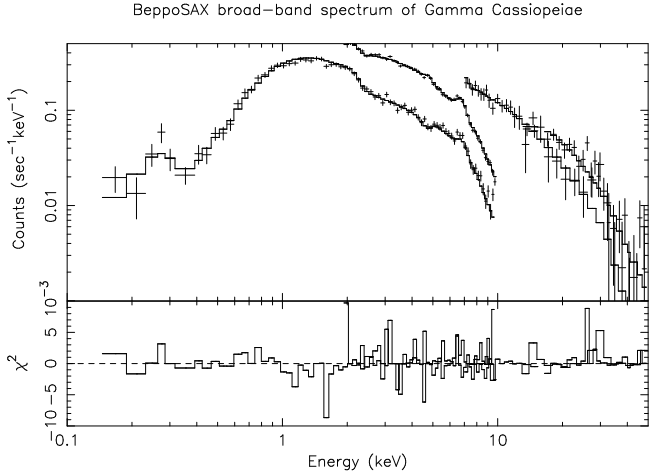}}
\caption[]{The BeppoSAX spectrum of $\gamma$-Cas measured by the NFI. 
The solid lines show the best-fit 2 component optically-thin thermal plasma 
(MEKAL) model (see text), folded through the instrumental responses. The 
contribution to $\chi^2$ are shown in the lower panel}
\end{figure}

\begin{figure}
\centerline{\psfig{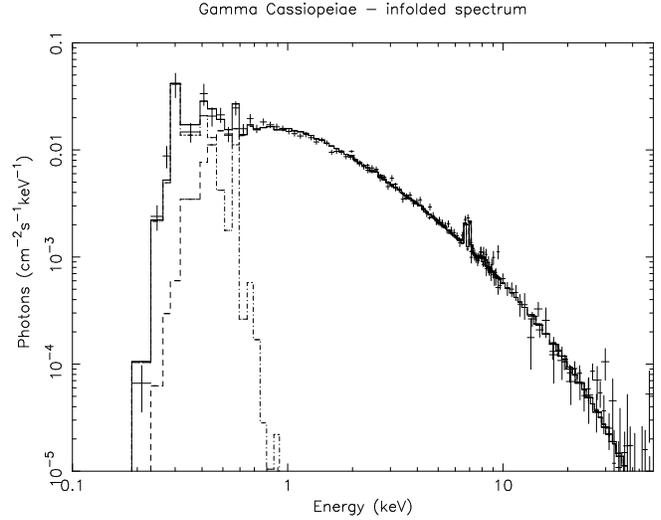}}
\caption[]{The best-fit incident photon spectrum of $\gamma$--Cas based on an absorbed two
temperature MEKAL model. The individual contributions are shown separately as long and short dashed lines }
\end{figure}

\begin{table}[htb]
\caption{\protect \small Best-fit spectral parameters.  Uncertainties are 
given at the 90\% confidence level for one interesting parameter. Line energies, 
widths and temperatures are in units of keV. Line fluxes and model fluxes are 
given in units of photon~cm$^{-2}$ s$^{-1}$ and 10$^{-10}$ erg 
cm$^{-2}$ s$^{-1}$, respectively. Column densities, ${\rm N_H}$, are in units of 
10$^{21}$~atom cm$^{-2}$ and equivalent widths, ${\rm EW}$, in units of eV} 
\label{lfit}
 
\begin{tabular}{ll} \hline
\noalign{\smallskip}
Parameter &  Value    \\
\multicolumn{2}{c}{Model 1: Absorbed power--law plus Fe line  \hfil  } \\ \hline
 
N${\rm _H}$                                & 2.75$\pm$0.20\\
$\alpha$                                   & 1.69$\pm ^{0.03} _{0.02}$    \\
E${\rm _{line}}$                           & 6.77${\pm 0.05}$  \\    
$\sigma{\rm _{line}}$                      & 0.21$\pm$0.07      \\       
EW                                         & 296$\pm$48   \\
Flux${\rm _{line}}$                        & (3.8$\pm$0.6)$\times$10$^{-4}$   \\
Model flux (2--10 keV)                     & 1.29            \\      
$\chi^2$/dof                               & 505/200              \\
\hline
\noalign{\smallskip}
\multicolumn{2}{c}{Model 2: Absorbed MEKAL          \hfil         } \\ \hline
N${\rm _H}$                                 & 1.55$\pm$0.09     \\
kT                                          & 12.5$\pm$0.4    \\
Abundances			            & 0.43$\pm$0.05\\
Model flux (2--10 keV)                      & 1.33 \\
$\chi^2$/dof                                & 266/202              \\
\hline
\noalign{\smallskip}
\multicolumn{2}{c}{Model 3: Absorbed MEKAL plus C line       \hfil     } \\ \hline
N$_H$                                       & 1.63$\pm$0.1      \\
kT                                          & 12.4$\pm$0.6    \\
Abundances			            & 0.42$\pm$0.05 \\
Flux${\rm _{line}}$                         & 0.35$\pm_{~0.14}^{~0.25}$ \\
$\sigma{\rm _{line}}$                       & $<$0.02      \\
EW                                          & 3220$\pm$1800   \\
Model flux (2--10 keV)                      & 1.33 \\
$\chi^2$/dof                                & 243/199              \\
\hline
\noalign{\smallskip}
\multicolumn{2}{c}{Model 4: Absorbed 2 component MEKAL \hfil } \\ \hline
N${\rm _H}$                                 & 1.7$\pm$0.15       \\
kT$_1$                                      & 12.3$\pm$0.6  \\
kT$_2$                                      & 0.05$\pm$0.01  \\
Abundances			            & 0.42$\pm$0.05\\
Model flux (2--10 keV)                      & 1.33   \\               
$\chi^2$/dof                                & 240/200              \\
\hline
\end{tabular}
\end{table}

\section{Discussion}
The comparison of BeppoSAX results with previous results 
can be misleading in view of the limited energy range of early
missions -- meaning that the results can be critically dependent 
on assumed emission models which may be inappropriate 
for the wide bandwidth of BeppoSAX. For example, ASCA data are 
perfectly consistent with a power--law continuum whereas BeppoSAX 
data are not, when energies above the ASCA upper energy threshold 
are taken into account. Globally, the BeppoSAX data show that the 
$\gamma$--Cas spectrum is consistent with an optically thin thermal 
plasma distribution which does not require non-thermal components 
-- as might be expected for accreting neutron star models.  The Fe 
line is a persistent feature 
of this source and is generally attributed to a blend of Fe XXV (6.7~keV) 
line emission and Fe XXVI (6.97~keV) emission produced in a highly ionized, 
optically thin thermal plasma. The implied temperature of $\sim$12~keV 
is perfectly consistent with that derived independently for the 
continuum. Such emission is most consistent with white dwarf scenarios. 
High mass neutron star systems, on the other hand, generally have strong emission 
at 6.4~keV and only weak, if any, emission at 6.7~keV. Also, 
such systems generally have non-thermal spectra which are well
described by a power-law distribution with a high energy cut-off and
may be expected to produce cyclotron line emission above $\sim$10~keV. 
None of these are observed by BeppoSAX. 
 
Until recently, the main problem with degenerate dwarf models has been 
reproducing the relatively high luminosity. However, this has been 
substantially relaxed in view of the recently revised source distance 
of 188~pc (ESA 1997) as opposed to the previously assumed value of 
250~pc. When coupled with a re-evaluation of wind parameters by Stee 
et al. (1995), the observed luminosity is no longer in conflict with 
current limits on the orbital motion. The lack of X-ray pulsations 
may indicate that the white dwarf does not possess a strong magnetic 
field or that the orbital period is long compared to the time--line of
the current X-ray database. The latter explanation would be consistent 
with the calculations of Kubo et al. (1998) who derive an orbital period 
for the system of 150~days. 
Lastly, in a critical review of X-ray emission models, Kubo et al. (1998) 
conclude there is no compelling evidence for the coronal model of Smith et al. 
(1998) since the very properties they claim are characteristic of coronal 
emission are also seen in the dwarf nova SS Cyg (Watson et al. 1985).

In summary, our results confirm previous ASCA and ROSAT measurements
and support the view of Kubo et al. (1998) that the characteristics 
of the X-ray emission are fully consistent with the conventional picture 
of $\gamma$--Cas as a binary system containing an accreting non-magnetic 
white dwarf, rather than neutron star binary or coronal emission models. 


\begin{acknowledgements}
The \sax\ satellite is a joint Italian and Dutch programme. TO 
acknowledges an ESA Research Fellowship. We thank the staff of 
the BeppoSAX Science Data Center for help with these observations 
and Gian Paolo Tozzi for useful comments. 
We also thank H. Barwig and C. Ries of the Universit\"ats-Sternwarte M\"unchen,
Observatorium Wendelstein, for their positive and rapid response to our
request for parallel observations. The Hipparcos catalogue
is available on line, at http://astro.estec.esa.nl/Hipparcos/catalog.html.
The quick--look RXTE ASM data were provided by the ASM/RXTE team 
(http://space.mit.edu/XTE/asmlc). This research has made use of the 
AFOEV database, operated at CDS, France (http://astro.u-strasbg.fr/afoev).

\end{acknowledgements}


\begin{thebibliography}{}
\bibitem[1976]{}

Association Francaise des Observateurs d'Etoiles Variables, 1998,
http://astro.u-strasbg.fr/afoev
\bibitem[1976]{}

Boella G., Butler R.C., Perola G.C., et al., 1997a, A\&AS 122, 299
\bibitem[1976]{}

Boella G., Chiappetti L., Conti G., et al., 1997b, A\&AS 122, 327
\bibitem[1976]{}



Doazan V., Franco M., Rusconi L., et al., 1983, A\&A 128, 171
\bibitem[1976]{}


ESA 1997, The Hipparcos and Tycho Catalogues, ESA SP-1200
\bibitem[1976]{}

 
Frontera F., Dal Fiume D., Robba N.R., et al. 1987, ApJ 329, L127
\bibitem[1976]{}
 
Haberl F., 1995, A\&A 296, 685
\bibitem[1976]{}
 
Horaguchi T., Kogure T., Hirata R.,  et al., 1994, PASJ 46, 9
\bibitem[1976]{}
 
Kaastra J.S., 1992, An X-Ray Spectral Code for Optically Thin Plasmas,
Internal SRON-Leiden Report, updated version 2.0
\bibitem[1976]{}

King A.R., Shaviv G., 1984, MNRAS  211, 883
\bibitem[1976]{}

Kubo S., Murakami T., Ishida M., Corbet R.D.H., 1998, PASJ 50, 417
\bibitem[1976]{}

Manzo G., Giarrusso S., Santangelo A., et al., 1997, A\&AS 122, 341
\bibitem[1976]{}

Mason K.O., White N.E., Sanford P.W., 1976, Nature 260, 690
\bibitem[1976]{}

Mewe R., Lemen J.R., van den Oord G.H.J., 1986, A\&AS 65, 511
\bibitem[1976]{}


Mukai K., Shiokawa K., 1993, ApJ 418, 863
\bibitem[1976]{}
  
Murakami T., Koyama K., Inoue H., Agrawal P.C., 1986, ApJ  310, L31
\bibitem[1976]{}
 


Pallavicini R., Golub L., Rosner R., et al., 1981, ApJ 248, 279
\bibitem[1976]{}

Parmar A.N., Martin D.D.E., Bavdaz M., et al., 1997, A\&AS 122, 309
\bibitem[1976]{}
 
Quirrenbach A., et al., 1997, ApJ 479, 477
\bibitem[1976]{}

Rosen S., Mason K., Cordova F., 1988, MNRAS 231, 549
\bibitem[1976]{}


Smith M., Robinson R.D., Corbet R.H.D., 1998, ApJ 503, 877
\bibitem[1976]{}

Stee Ph., de Araujo F.X., Vakili F., et al., 1995, A\&A 300, 219
\bibitem[1976]{}

 
Vakili F., et al., 1994. In: Pulsation, Rotation and Mass
Loss in Early-Type Stars, (eds.) Balona L.A., Hendrichs H.F., 
Le Contel J. M., International Astronomical Union. Symposium
No. 162 held at Juan-les-Pines, France, p. 241
\bibitem[1976]{}

 

Watson M.G., King A.R., Heise J. 1985, Space Sci. Rev. 40, 127
\bibitem[1976]{}

 
\end{thebibliography}
\end{document}